\documentclass[doublecol]{epl2}

\newcommand{\be}{\begin{equation}}
\newcommand{\ee}{\end{equation}}
\newcommand{\beq}{\begin{eqnarray}}
\newcommand{\eeq}{\end{eqnarray}}

\usepackage{amsmath}
\usepackage{amssymb}
\usepackage{graphicx}
\usepackage{xcolor}
\usepackage{bm}
\usepackage{relsize}

\title{Wetting of soft substrates}
\shorttitle{Wetting of soft substrates} 

\author{M. Napi\'orkowski\inst{1} \and L. Schimmele\inst{2} \and S. Dietrich\inst{2,3}}
\shortauthor{M. Napi\'orkowski \etal}

\institute{                    
  \inst{1} Institute of Theoretical Physics, Faculty of Physics, University of Warsaw, Pasteura 5, 02-093 Warszawa, Poland\\
  \inst{2} Max-Planck-Institut f\"ur Intelligente Systeme, Heisenbergstr. 3, D-70569 Stuttgart, Germany\\
  \inst{3} IV. Institut f\"ur Theoretische Physik, Universit\"at Stuttgart, Pfaffenwaldring 57, D-70569 Stuttgart, Germany}

\abstract{Within mean-field theory we study wetting of elastic substrates. Our analysis is based on a grand canonical free energy functional of the  
fluid number density and of the substrate displacement field. The substrate is described in terms of the linear theory of elasticity,
parametrized by two Lam\'e coefficients. 
The fluid contribution is of the van der Waals type. Two potentials characterize the interparticle interactions in the system. The long-ranged
attraction between the fluid particles is described by a potential  
$w(\bm{r})$, and $v(\bm{r})$ characterizes the substrate-fluid interaction.  
By integrating out the elastic degrees of freedom we obtain an effective theory for the fluid number density alone. Its structure is similar to the one for 
wetting of an inert substrate. However, the potential $w(\bm{r})$ is replaced by an effective potential
which, in addition to $w(\bm{r})$, contains a term bilinear in $v(\bm{r})$.  
We discuss the corresponding wetting transitions in terms of an effective interface potential $\omega(\ell)$, where $\ell$ denotes the thickness 
of the wetting layer. 
We show that in the case of algebraically decaying interactions the elasticity of the substrate may suppress critical wetting transitions, 
and may even turn them first order.}

\begin{document}

\maketitle

\section{1. Introduction}
The past decades have produced a wealth of experimental and theoretical results concerning wetting (see the reviews in Refs. 
\cite{PdG1985,DC1986,SD1988,FLN1991,HBS2008,RD2008,BEIMR2009,DRN2013}). Various aspects of systems, in which wetting occurs, have been examined. 
They include details of the internal structure of the wetting fluids as well as the structures of substrates, both geometrical and chemical.  
These structural attributes of both wetting fluids and substrates are displayed on a macroscopic scale as they influence the order and the location of 
wetting transitions. 
Usually, in the theoretical description of such systems the substrate, whether structured or not, 
has been considered as an inert spectator phase, i.e., it serves as the source of an external force acting  on the fluid but itself it is not influenced 
by the changes in the state of the fluid. Only rarely feedback mechanisms have been incorporated into the theoretical analysis which would allow one to 
overcome the limitations of the concept of an inert substrate. This idealization 
has been relaxed in recent studies of adsorption phenomena and morphological transitions of adsorbed phases and, in particular, elastic 
substrates have been considered \cite{BBD2010,BD2012,MDSA2012,SCPWJHGPWWD2013,LJ2013,MMPFAP2013,VMCPLRC2013,LWBDAS2014,JF2014}. Examples of such particularly soft substrates comprise gels and rubbers. \\

Here we consider a simple liquid wetting an elastic substrate and address the issue how the elasticity of the substrate influences the wetting transition.  
Our mean-field analysis is based on the grand canonical free energy functional $\Omega([\rho], [{\bm{u}}]; T, \mu)$, where $\rho(\bm{r})$ 
is the fluid number density and  $\bm{u}(\bm{r})$ is the displacement field of the elastic substrate.  The functional is a function of temperature $T$ and of the chemical potential $\mu$ of the fluid. The details of the model are described in Sec. 2 where the equations for the equilibrium number density $\overline{\rho}(\bm{r})$ and for the displacement field 
$\overline{\bm{u}}(\bm{r})$ are derived and analyzed. This section 
contains also the derivation of the effective theory, which results from integrating out the displacement field. It depends exclusively on the fluid number 
density and employs the concept of an effective interaction between the fluid particles. In Sec. 3  this effective theory serves as the starting 
point for a 
coarse-graining process in which the positions of the liquid-gas interface and of the solid-liquid interface are left as the only relevant degrees 
of freedom.  The effective interface potential, describing the effective interaction of the liquid-gas interface with the elastic substrate, depends on
these two degrees of freedom. The general form of the effective interface potential is derived and compared with its form in the case of an inert 
(nonelastic) substrate. The knowledge of the effective interface potential allows for a rather straightforward analysis of wetting transitions. 
In Sec. 4 we restrict our analysis of wetting to the case of long-ranged, algebraically decaying forces, and discuss the influence of 
substrate elasticity on the order and the location of the wetting transition as compared with the inert case. Section 5 contains a brief summary 
of our results. \\

\section{2. Model and derivation of the effective interaction among the fluid particles as induced by an elastic substrate}
Our model system consists of a simple fluid with number density $\rho(\bm{r})$ and an isotropic elastic substrate characterized by 
the displacement field $\bm{u}(\bm{r})$. 
The elastic substrate and the fluid occupy regions with volume $V_{s}$ and $V_{f}$, respectively.  The equilibrium state of this system 
minimizes the following grand canonical functional of the fields $\rho(\bm{r})$ and $\bm{u}(\bm{r})$:
\beq
\label{omega1}
\Omega([\rho],[\bm{u}]; T,\mu, V_{s}, V_{f}) =   \nonumber \hspace{2cm} \\
  \Omega_{f}([\rho];T,\mu,V_{f}) + \Omega_{s}([\bm{u}]; T, V_{s}) + 
\delta\Omega([\bm{u}],[\rho];V_{s},V_{f}) \,.
\eeq
The mean-field free energy $\Omega_{f}$ of the fluid   has the following form  \cite{DN01}:
\beq
\Omega_{f}([\rho]; T, \mu,V_{f}) =  \int\limits_{V_{\mathlarger{f}}} d^3r f_{h}(\rho(\bm{r})) - \, \mu \int\limits_{V_{\mathlarger{f}}} d^3r \rho(\bm{r}) \, + 
\nonumber \\ \frac{1}{2}  \int\limits_{V_{\mathlarger{f}}} d^3r\,' \, \int\limits_{V_{\mathlarger{f}}} d^3r\,'' \rho(\bm{r}\,') \rho(\bm{r}\,'') w(\bm{r}\,' - \bm{r}\,'') ,
 \eeq
 where $f_{h}(\rho(\bm{r}))$ is the free energy density of the reference fluid with short-ranged hard repulsion only, and  where 
 $w(\bm{r}\,'- \bm{r}\,'')$ is the attractive part of the spherically symmetric interaction potential among the fluid particles. 
 The elastic free energy $\Omega_{s}$ 
of the substrate \cite{LLElastic} is 
\beq
\label{els3}
\Omega_{s}([\bm{u}]; T, V_{s}) =  \nonumber \hspace{2cm} \\
\int\limits_{V_{\mathlarger{s}}} d^3r \left[ \frac{\lambda}{2} \,\epsilon_{ii}(\bm{r})\epsilon_{kk}(\bm{r})
           + \nu\,\epsilon_{ik}(\bm{r}) \epsilon_{ki}(\bm{r})\right] ,
\eeq
where 
\beq
\epsilon_{ij}(\bm{r}) = \frac{1}{2} \left(\frac{\partial u_{i}(\bm{r})}{\partial x_{j}} + \frac{\partial u_{j}(\bm{r})}{\partial x_{i}} \right) 
\eeq
is the strain tensor; $\lambda$ and $\nu$ are the Lam\'e coefficients (summation over repeated indices is implicit).
Equation~(\ref{els3}) suppresses the contribution $\Omega_{s}([\bm{u}=0]; T, V_{s})$, which is independent of $\rho$ and thus
irrelevant for our present purposes.  
The fluid-substrate interaction $\delta\Omega$ has the following form \cite{DN01}: 
\beq
\delta\Omega([\bm{u}],[\rho],V_{s},V_{f}) =  \nonumber \hspace{2cm} \\ \int\limits_{V_{\mathlarger{s}}}d^3r\,' \int\limits_{V_{\mathlarger{f}}}d^3r\,'' \rho_{s}(\bm{r}\,') \rho(\bm{r}\,'') v(\bm{r}\,' - \bm{r}\,'') , 
\eeq
where 
$v(\bm{r}\,'- \bm{r}\,'') $  is the spherically symmetric interaction potential between a fluid and a substrate particle and $\rho_{s}(\bm{r})$ denotes 
the number density of the substrate particles. It is related to the displacement field $\bm{u}(\bm{r})$ via  
\beq
\label{rs01}
\rho_{s}(\bm{r}) = \rho_{0} \,\left[1- \mathrm{div}  \,\bm{u}(\bm{r}) \right] \quad,
\eeq
where $\rho_{0}$ is the spatially constant number density of the substrate without displacements. 
The equilibrium number density $\overline{\rho}(\bm{r})$ and the displacement field $\overline{\bm{u}}(\bm{r})$ minimize the functional  
$\Omega([\rho],[\bm{u}]; T,\mu, V_{s}, V_{f})$ given in Eq.~(\ref{omega1}) and solve the equations 
\beq
\label{eq1}
f_{h}'(\overline{\rho}(\bm{r})) + \int\limits_{V_{\mathlarger{f}}} d^3r\,' \overline{\rho}(\bm{r}\,') \, w(\bm{r}- \bm{r}\,') +  \nonumber \\
           \rho_{0}\int\limits_{V_{\mathlarger{s}}} d^3r\,' v(\bm{r}- \bm{r}\,')  (1- \mathrm{div} \, \overline{\bm{u}}(\bm{r}\,')) = \mu
\eeq
and
\beq
\label{eq2}
\nu\, \Delta \overline{\bm{u}}(\bm{r}) + (\nu + \lambda)\, \mathrm{grad \, div}  \, \overline{\bm{u}}(\bm{r}) - \nonumber \\
          \rho_{0} \,\mathrm{grad} \,\int\limits_{V_{\mathlarger{f}}} \,d^3r\,' \overline{\rho}(\bm{r}\,') \, v(\bm{r}-\bm{r}\,') = 0 . 
\eeq
The boundary condition for the displacement field is $\overline{u}_{z}(x,y,z=0) = h(x,y)$, where $z=h(x,y)$ denotes the position of the 
(in general non-planar) solid-liquid interface. The function $h(x,y)$ follows from the balance of forces at the substrate - fluid interface. 
In order to avoid a clumsy notation, we omit the overbars so that from here on $\rho(\bm{r})$ and $\bm{u}(\bm{r})$ denote the 
equilibrium number density and displacement fields. \\

In the following we consider systems which are spatially homogeneous and of macroscopic lateral extent in the $x$ and $y$ directions.  
The substrate extends vertically to $-L_{s} \le z \le 0$ with 
$L_{s}>0$.  In such a case one has $h(x,y)=h=const$ and $\rho(\bm{r}) = \rho(z)$, $\bm{u}(\bm{r}) = (u_{x}(z), u_{y}(z), u_{z}(z))$, 
and  Eq.\,(\ref{eq2}) takes the following form: 
\beq
\label{u01}
    \gamma \, \frac{d^2 u_{z}}{dz^2} = \rho_{0} \int\limits_{h}^{\infty} dz' \,\rho(z') \,\frac{\partial \widetilde{v}(z-z')}{\partial z} \quad,
\eeq
with $\gamma \equiv \lambda + 2 \nu$, 
\beq
\label{v01}
\widetilde{v}(z) = \int\limits_{\mathbb{R}} dx \int\limits_{\mathbb{R}} dy \, v(\bm{r}) \quad, 
\eeq  
and 
\beq
\label{u011}
\frac{d^2 u_{x}}{d^2 z} = \frac{d^2 u_{y}}{d^2 z} = 0 \quad.
\eeq
Equation (\ref{u011}) has as solutions $u_{x}(z)=u_{y}(z)=0$, while integrating Eq.~(\ref{u01}) once gives 
\beq
\label{u001}
\frac{du_{z}(z)}{dz} = \frac{\rho_{0}}{\gamma} \int\limits_{-L_{\mathlarger{s}}}^zdz' \int\limits_{h}^{\infty}dz''\rho(z'') 
\frac{\partial\tilde{v}(z'-z'')}{\partial z'} + C 
\eeq
with integration constant $C$. At $z=0$ one has the force balance equation 
\beq
\label{bal01}\gamma \left.\frac{du_{z}(z)}{dz} \right|_{z=0} = \overline{\sigma}_{zz}\, ,
\eeq
where $\overline{\sigma}_{zz}$ is the stress tensor evaluated at $z=0$; $-\overline{\sigma}_{zz}$ accounts for those forces,  exerted by the fluid on the substrate, which are modeled as contact forces, i.e., acting at
the surface only. These contact forces plus the sum of the body forces per area (the latter are displayed  explicitly on the rhs of Eqs. (\ref{eq2}) and  (\ref{u01})) 
must be equal to the pressure $p$ in the fluid in order to achieve a thermodynamic consistent modeling. Using approximate fluid number-density
profiles would also lead to violations of thermodynamic consistency if not enforced by a proper choice of $\overline{\sigma}_{zz}$.
A thermodynamic consistent modeling ensures a proper asymptotic displacement field.   
The boundary condition in Eq. (\ref{bal01}) determines the integration constant $C$ in Eq. (\ref{u001}):   
\beq
\label{C01}
-\gamma C=-\overline{\sigma}_{zz}+\rho_{0}\int\limits_{-L_{\mathlarger{s}}}^0dz'\int\limits_{h}^{\infty}dz''\rho(z'')\frac{\partial \tilde{v}(z'-z'')}{\partial z'}=p,
\eeq  
where the last equality results from the above mentioned requirement of thermodynamic consistency. This relation gives 
$C = - p/\gamma$.
Integration of Eq.~(\ref{u001}), combined with the boundary condition $u_{z}(0)=h$, yields 
\beq
\label{hLs01}
\frac{h}{L_{\mathlarger{s}}} =-\frac{1}{\gamma}\left[p+\frac{\rho_{0}}{L_{\mathlarger{s}}}\int\limits_{-L_{\mathlarger{s}}}^0dz'\int\limits_{h}^{\infty}dz''z'\rho(z'') 
\frac{\partial\tilde{v}(z'-z'')}{\partial z'}\right] .
\eeq
In the limit of large $L_{s}$  the above equation reduces to $h/L_{s} = - p/\gamma + O(L_{s}^{-2})$ which is in line with the elasticity theory of unilateral compression \cite{LLElastic}. After inserting the  above results into Eq. (\ref{eq1}) and shifting the reference position of the substrate-liquid interface from $z=h$ to $z=0$ (this shift does not influence the thickness of the wetting layer), 
in the limit $L_{s} \rightarrow \infty$ one obtains the following equation for the fluid number density $\rho(z)$: 
\be
\label{eq3302}
\mu = f_{h}'(\rho(z)) + \rho_{0}\left(1+\frac{p}{\gamma}\right)\int\limits_{-\infty}^{0}dz'\widetilde{v}(z -z') \, + 
\ee
\be
\int\limits_{0}^{\infty}dz'\rho(z') \left[\widetilde{w}(z-z')-\frac{\rho_{0}^2}{\gamma}\int\limits_{-\infty}^{0}dz''\widetilde{v}(z-z'')\widetilde{v}(z''-z')\right],  \nonumber 
\ee
where $\widetilde{w}(z)=\int\limits_{\Bbb{R}}dx \int\limits_{\Bbb{R}}dy \, w(\bm{r})$. 
This equation has the same structure as in the case of an inert substrate with a spatially constant number density $\rho_{s}(\bm{r})=\rho_{0}$ \cite{DN01} 
(see Eq. (\ref{eq1}) for $\bm{u} \equiv 0$) except for two essential modifications: (i) 
in the second term on the rhs the "renormalized" number density $\rho_{0}(1+ p/\gamma)$ of substrate particles replaces the 
constant density $\rho_{0}$ of an inert substrate; 
(ii) in the third term on the rhs there is an effective interparticle potential $\widetilde{w}_{eff}(z,z')$ (in square brackets) which replaces
the interparticle potential $\widetilde{w}(z-z')$ (see the second term on the lhs of Eq. (\ref{eq1})) such that 
\beq
\label{weff01}
\widetilde{w}_{eff}(z,z') =  \widetilde{w}(z-z') \, + \, \widetilde{w}_{elastic}(z,z')
\eeq
where 
\beq
\label{welas01}
\widetilde{w}_{elastic}(z,z') \, = \, - \frac{\rho_{0}^2}{\gamma} \int\limits_{-\infty}^{0} dz'' \widetilde{v}(z-z'')\,\widetilde{v}(z''-z') .
\eeq
In the expression for the effective interparticle potential $\widetilde{w}_{eff}(z,z')$ (Eq.~(\ref{weff01})) the additional term  $\widetilde{w}_{elastic}(z,z')$
(Eq.~(\ref{welas01})) 
stems from the deformation of the elastic substrate caused by its interaction with 
fluid particles \cite{LK01}. It is represented by a term bilinear  in the substrate potential $v(z)$ and is absent for the inert (nonelastic) substrate. 
This extra term is symmetric in its arguments. However, contrary to $\widetilde{w}(z-z')$, it is not translationally invariant in the vertical direction. 
For $z=z'$ this term is negative. 

\section{3. The effective interface potential}
In this section we discuss wetting of an elastic substrate based on the  concept of the effective interface potential $\omega(\ell)$, 
where $\ell$ denotes the imposed  thickness of the wetting layer. The equilibrium thickness $\overline{\ell}$ of the wetting layer 
minimizes $\omega(\ell)$. In order to derive an approximate, but surprisingly reliable, expression for 
$\omega(\ell)$, we consider a piece-wise constant fluid number density profile \cite{DN01} 
 \begin{equation}
 \label{sk001}
 \rho(z) = \left\{
 \begin{array}{ccl}
 \rho_{\ell} & for & 0 < z < \ell \\ 
  \rho_{g} & for & \ell < z \leq L \\
  0        & otherwise & 
  \end{array}
  \right. ,
  \end{equation} 
where $\rho_{\ell}$ and $\rho_{g}$  denote the bulk liquid and gas number densities, respectively. In the spirit of the thermodynamic limit the cut-off $L$ 
is introduced in order to avoid infinite 
expressions. (After deriving the relevant expression for the effective interface potential we take the limit $L=\infty$.) After inserting the above 
expression into Eq.~(\ref{omega1}) one  identifies the bulk $\Omega_{b} = V_{f} \varphi(\rho_{g})$ and surface $\Omega_{s}$ contributions 
to $\Omega([\rho],[\bm{u}]; T,\mu, V_{s}, V_{f})$. The surface contribution per area $A$ takes the form
 \beq
  \label{omegaellh}
  \lim\limits_{A \rightarrow \infty} \,\frac{\Omega_{s}}{A}  =  \sigma_{\ell g} + \widetilde{\sigma}_{s \ell} + \sigma_{g v} + \widetilde{\omega}(\ell) \quad, 
  \eeq
 where  $\sigma_{\ell g}$ is the liquid-gas surface tension coefficient, $\widetilde{\sigma}_{s \ell}$ is the coefficient of the substrate-liquid surface tension, 
 $\sigma_{g v}$ is the coefficient of the surface tension associated with the gas-vacuum interface (introduced by the cut-off at $z=L$), and $\widetilde{\omega}(\ell)$ is the effective liquid-gas interface potential.  
 The above quantities are defined as follows: 
  \beq
  \label{w0}
  \varphi(\rho)=f_{h}(\rho)+\frac{1}{2} w_{0} \rho^2 - \, \mu \rho = -p_{h}(\rho) - \frac{1}{2} w_{0} \rho^2 , 
  \eeq
 where $w_{0}=\int\limits_{\Bbb{R}^3} d^3r \,w(\bm{r})  < 0$  and $p_{h}(\rho) = - f_{h}(\rho) - w_{0} \rho^2 + \mu \rho$ is the pressure of the 
 reference fluid. Within the above sharp-kink approximation the coefficient of the liquid-gas surface tension is  
   \beq 
  \label{sigmalg}
  \sigma_{\ell g} = - \frac{(\Delta \rho) ^2}{2}  \int\limits_{0}^{\infty} dz t(z) \, ,
  \eeq
  where $\Delta \rho = \rho_{\ell}-\rho_{g}$ and 
  \beq
  \label{tz}
  t(z) = \int\limits_{z}^{\infty}dz' \widetilde{w}(z')\,\,,\,\, s(z) = \int\limits_{z}^{\infty}dz' \widetilde{v}(z')\,.  
  \eeq
  The coefficient of the substrate-liquid surface tension is 
  \beq
  \label{tsigmawl}
  \widetilde{\sigma}_{s \ell} = \sigma_{s \ell} + 
          \frac{\rho_{0}\rho_{\ell}}{\lambda + 2\nu} \left( p\int\limits_{0}^{\infty} dz s(z) - 
  \frac{\rho_{0}\rho_{\ell}}{2} \int\limits_{0}^{\infty} dz s^2(z) \right)
  \eeq
  where 
  \beq
  \label{sigmawl}
  \sigma_{s \ell} = \rho_{0}\rho_{\ell} \int\limits_{0}^{\infty} dz \,s(z) - \frac{\rho_{\ell}^2}{2} \int\limits_{0}^{\infty} dz \,t(z)  
  \eeq
  corresponds to the limiting case of an inert (nonelastic) substrate. The effective interface potential takes the form  
  \beq
  \label{tomegal}
  \widetilde{\omega}(\ell) = \omega(\ell) + \frac{\rho_{0}^2 \Delta\rho}{\gamma} \left[ \rho_{\ell} \int\limits_{0}^{\infty} s(z) 
  s(\ell+z) \right. \nonumber \\ \left. -\frac{\Delta\rho}{2} \int\limits_{\mathlarger{\ell}}^{\infty} dz \, s^2(z) - \frac{p}{\rho_{0}}\int\limits_{\mathlarger{\ell}}^{\infty} dz s(z)  \right] , 
	\eeq 
  where
  \beq
  \label{omegal}
  \omega(\ell)= \Delta \rho \,\left\{\rho_{\ell} \int\limits_{\mathlarger{\ell}}^{\infty} dz\, t(z) - \rho_{0} \int\limits_{\mathlarger{\ell}}^{\infty} dz \,s(z) \right\} + \ell \Delta\varphi 
  \eeq
is the effective interface potential in the inert (nonelastic) substrate case, and $\Delta\varphi = \varphi(\rho_{\ell}) - \varphi(\rho_{g})$. 
An alternative way of deriving the expression in Eq.~({\ref{tomegal}) is to start with the expressions for an inert substrate (Eqs. (\ref{sigmawl}) and (\ref{omegal})) for the 
solid--liquid interfacial tension and the effective interface potential, respectively, and to replace then the number density $\rho_{0}$ by 
the "renormalized" number density $\rho_{0}(1+ p/\gamma)$, and 
the interparticle potential $\widetilde{w}(z-z')$ in Eq.~(\ref{tz}) by the effective interparticle potential 
$\widetilde{w}_{eff}(z,z')$ defined in 
Eq.~(\ref{weff01}). Collecting the $\ell$ dependent terms gives $\widetilde{\omega}(\ell)$.   
We note that the surface free energy density $\Omega_{s}/A$ has the same structure as in the previously studied case of wetting of an {\it inert} substrate,
except that in the present elastic case there are two modifications: (i) The inert substrate-liquid surface tension coefficient $\sigma_{s \ell}$ 
in Eq.~(\ref{sigmawl}) is replaced by 
  $\widetilde{\sigma}_{s \ell}$ given in Eq.~(\ref{tsigmawl}). The presence of an elastic substrate is indicated by two extra terms, one linear and one bilinear 
in the substrate potential $v$ (Eqs. (\ref{tsigmawl}) and (\ref{sigmawl})). The coefficients in front of these terms  depend on the substrate 
elastic constants via $\gamma$, on the fluid pressure $p$, and on the number densities $\rho_{0}$ and $\rho_{\ell}$. 
(ii) The  effective interface potential -- which in the case of an {\it inert} substrate takes the form  $\omega(\ell)$ in Eq. (\ref{omegal}) -- is, 
in the elastic case, replaced  by $\widetilde{\omega}(\ell)$ in Eq.~(\ref{tomegal}) which -- in addition to $\omega(\ell)$ -- 
contains extra terms depending on the substrate potential $v$. They are linear and bilinear in  $v$, respectively and additionally depend on the substrate elastic constants.  
The equilibrium wetting layer thickness $\overline{\ell}$ minimizes $\widetilde{\omega}(\ell)$ so that it fulfills 
\beq
\label{eep1}
\rho_{0} \,\left(1+\frac{p}{\gamma}\right)\,u(\overline{\ell})\, - \rho_{\ell}\, t(\overline{\ell}) + \frac{\Delta \varphi}{\Delta \rho} + \nonumber \\ 
+ \,\frac{\rho_{0}^2}{\gamma} \left\{\frac{\Delta\rho}{2} \,u^2(\overline{\ell})  + \rho_{\ell} \int\limits_{0}^{\infty}dz\, u(z)\,u'(\overline{\ell}+z) \right\}  \,= \,0 .
\eeq
For thermodynamic states $(T,\mu)$ close to gas-liquid coexistence $(T,\mu_{0}(T))$ the ratio $\Delta\varphi/\Delta\rho$ measures the deviation
from two-phase coexistence: 
 $\Delta\varphi/\Delta\rho \approx \mu - \mu_{0}(T)$. In particular, 
first-order and critical wetting transitions occur at gas-liquid coexistence, whereas complete wetting occurs off coexistence;
prewetting occurs at and off two-phase coexistence.  

\section{4. Effective interface potential for long-ranged forces}
In this section we focus our analysis on systems with potentials $w(\bm{r})$ and $v(\bm{r})$ which, at large distances,  
describe long-ranged attraction with an algebraic decay $\sim r^{-6}$.  As a result, at large distances,  the functions $t(z)$ and $s(z)$ also decay algebraically: 
\beq
t(z) = - \sum\limits_{k \geq 3} \,  \frac{t_{k}}{z^k} \,\,,\,\,
s(z) = - \sum\limits_{k \geq 3} \,  \frac{s_{k}}{z^k} 
\eeq
with $t_{3}, s_{3} > 0$.  In this case the effective interface potential $\widetilde{\omega}(\ell)$ has, for large wetting layer thicknesses $\ell$, the following form 
\cite{DN01}:
\beq
\label{wcoef01}
\widetilde{\omega}(\ell) = \frac{\widetilde{\omega}_{2}}{\ell^2} \, + \, \frac{\widetilde{\omega}_{3}}{\ell^3} \,  + \,   \frac{\widetilde{\omega}_{4}}{\ell^4} \,  
+ \, O(\ell^{-5}\ln\ell) \quad,
\eeq
where 
\beq
\label{omt2}
\widetilde{\omega}_{2} = \omega_{2} \, + \,\frac{p}{\gamma} \, \frac{\rho_{0} \,\Delta\rho \,s_{3}}{2} \quad,
\eeq
\beq
\label{omt3}
\widetilde{\omega}_{3} = \omega_{3} \, + \,\frac{p}{\gamma} \, \frac{\rho_{0} \,\Delta\rho \,s_{4}}{3} - \, s_{3}\,\frac{\rho_{0}^2 \rho_{\ell} \Delta\rho}{\gamma}\,\int\limits_{0}^{\infty} dz \,s(z) \quad,
\eeq
and 
\beq
\label{omt4}
\widetilde{\omega}_{4} = \omega_{4} \, + \,\frac{p}{\gamma} \, \frac{\rho_{0} \,\Delta\rho \,s_{5}}{4} - \, s_{4}\,\frac{\rho_{0}^2 \rho_{\ell} \Delta\rho}{\gamma}\,\int\limits_{0}^{\infty} dz \,s(z) \nonumber \\ \, + \, 3\,s_{3} \,\frac{\rho_{0}^2 \rho_{\ell} \Delta\rho}{\gamma}\,\int\limits_{0}^{\infty} dz \,z\,s(z)      \quad. 
\eeq
In the case of an inert (nonelastic) substrate the effective interface potential has the analogous form 
\beq
\label{wcoef02}
\omega(\ell) = \frac{\omega_{2}}{\ell^2} \, + \, \frac{\omega_{3}}{\ell^3} \,  + \,   \frac{\omega_{4}}{\ell^4} \,  + \, O(\ell^{-5}\ln\ell) \quad,
\eeq
where the coefficients $\omega_{i}$, i=2,3,4, are given by 
\beq
\label{om2}
\omega_{i} = \frac{\Delta\rho}{i} \,\left(\rho_{0}\,s_{i+1} \, - \, \rho_{\ell}\,t_{i+1} \right) .   
\eeq
The elasticity induced modification of the effective interface potential, i.e., turning $\omega(\ell)$ into  $\widetilde{\omega}(\ell)$, may substantially alter the wetting scenario. We note that in the above formulae the substrate elasticity is taken into account via the Lam\'e coefficients $\lambda$ and $\nu$ which enter via the combination 
$\gamma=\lambda + 2\nu$ characteristic for unilateral compression \cite{LLElastic}. In particular, for $\gamma \rightarrow \infty$ one has 
$\widetilde{\omega}(\ell) \rightarrow \omega(\ell)$. The impact of substrate elasticity on wetting could be demonstrated most clearly by varying the coefficient $\gamma$  without a simultaneous, substantial modification of the substrate-fluid interaction potential $v$.
In particular, this possibility might be checked by numerical simulations of wetting on substrates with varying elastic properties. \\ 
In the following discussion we thus anticipate a situation in which the hypothetical modification of $\gamma$ takes place at practically unchanged values of 
the coefficients $s_{i}$, i=3,4,5. Critical wetting corresponds to 
$\mu=\mu_{0}(T)$ and to the temperature $T$ approaching the wetting temperature $\widetilde{T}_{w}$ from below. In this case  the coefficients 
$\widetilde{\omega}_{i}$  of the effective interface potential in Eq.~(\ref{wcoef01}) are such that $\widetilde{\omega}_{2}(T < \widetilde{T}_{w}) \leq 0$ and 
$\widetilde{\omega}_{3}(T = \widetilde{T}_{w}) > 0$. The critical wetting temperature $\widetilde{T}_{w}$ is given implicitly by the equation $\widetilde{\omega}_{2}(T = \widetilde{T}_{w}) = 0$. 
Similarly, in the case of critical wetting of an inert substrate one has $\omega_{2}(T < T_{w}) < 0$ and 
$\omega_{3}(T = T_{w}) > 0$ with the inert (nonelastic) substrate critical wetting temperature $T_{w}$ defined by the equation $\omega_{2}(T = T_{w})=0.$ In Eq.~(\ref{omt2}) 
the additional term $p\,\rho_{0} \,\Delta\rho \,s_{3}/2\gamma$, which is due to a contribution to the strain tensor describing
a homogeneous compression, is positive and thus 
$\widetilde{\omega}_{2}(T) > \omega_{2}(T)$. If the inert substrate exhibits critical wetting, i.e., $\omega_{2}(T \leq T_w) \leq 0$ 
and $\omega_{3}(T = T_w)>0$, in the 
corresponding case of an elastic substrate two scenarios are possible. First, if nonetheless $\widetilde{\omega}_{2}(T < \widetilde{T}_{w}) < 0$ 
and $\widetilde{\omega}_{3}(T = \widetilde{T}_{w}) > 0$, the elastic substrate undergoes citical wetting as well, but at a lower critical wetting temperature 
$\widetilde{T}_{w} < T_{w}$. The relative shift $(\widetilde{T}_{w} - T_{w})/T_{w}$ of the critical wetting temperature  can be determined  using Eqs. (\ref{omt2}) and (\ref{om2}) and is approximately given by 
\beq
\label{Twet}
\frac{\widetilde{T}_{w}-T_{w}}{T_{w}} \approx \frac{p_{0}(T_{w})}{\gamma(T_{w})} \, 
\left(\frac{d \rho_{\ell}}{d T} \frac{T}{\rho_{\ell}}\right)^{-1}_{T=T_{w}}, 
\eeq
where $p_{0}(T)$ denotes the liquid-gas coexistence pressure of the wetting fluid. The rhs of Eq.(\ref{Twet}) is dominated by the 
dimensionless ratio $p_{0}(T_{w})/\gamma(T_{w})$. The coefficient $\gamma$ of unilateral compression can be expressed in terms of Young's modulus $E$ and Poisson's ratio $\sigma$ as $\gamma = E (1-\sigma)/((1+\sigma)\,(1-2\sigma))$. In the case of, e.g., propanol wetting an appropriately selected silicone gel ($E=6$ kPa and $\sigma=0.46$) one has $p_{0}(T_{w})/\gamma(T_{w})$ = 0.09  \cite{DHMKC2002,GGLR2011,GG2011}. This factor vanishes in the limit of 
non-elastic substrates. The remaining factor in Eq. (\ref{Twet}) is of order unity. \\
Second, in a more exciting case the additional term in Eq.~(\ref{omt2}) may lead to a situation in which  $\widetilde{\omega}_{2}(T) > 0$
for all temperatures. This 
means that the elasticity of the substrate precludes critical wetting altogether although it does occur in the limiting inert (nonelastic) case. 
If  $\widetilde{\omega}_{2}(T) > 0$, there still exists the 
possibility that critical wetting of an inert substrate turns first order for elastic substrates. This happens if, together with  
 $\widetilde{\omega}_{2}(T) > 0$, one has $\widetilde{\omega}_{3}(T) < 0$ and $\widetilde{\omega}_{4}(T) > 0$. In this situation first-order wetting 
 takes place between two wetting layer thicknesses: large but finite and macroscopically large. On the other hand, within the above scenario, 
it follows from Eq.~(\ref{omt2}) 
 that the opposite situation of first-order wetting of an inert substrate turning critical for the elastic substrate is not possible. 
 
 \section{5. Summary and perspectives}
 We have considered a mean field theory of wetting of elastic substrates which is based on the minimization of a suitable free energy functional.
 This functional encompasses the contributions from the fluid, described by its number density, the elastic substrate described by its displacement field, and 
 their coupling. The equilibrium fluid number density and the substrate displacement fulfill a set of coupled integro-differential equations. 
 By integrating out the substrate degrees of freedom, we have obtained an effective theory for the fluid alone. It has the same structure as in the 
 case of adsorption on an inert substrate, but it is characterized by the effective interparticle interactions. 
 In a next step, we studied a reduced description 
 of the fluid density, according to which in the context of analyzing wetting the only relevant degree of freedom is the thickness of the adsorbed liquid layer.  
 We have derived an analytic expression for the effective interface potential which is a function of the thickness of the wetting layer and which allows 
 the discussion of various wetting scenarios. In particular, for the case of long-ranged van der Waals interparticle forces we point out 
 the situation in which the inert (nonelastic) substrate undergoes 
 critical wetting while the elasticity of the substrate either induces a
 decrease of the critical wetting temperature as compared to the case of wetting of an inert substrate, or precludes critical wetting 
 of the elastic substrate altogether. In the latter case there exists the possibility that critical wetting of an inert substrate 
 turns first order in the elastic case. However, the opposite scenario that a first-order wetting transition of an inert substrate 
 becomes critical for the elastic substrate is not possible. The above conclusions are 
 in line with the rather rare experimental evidence of critical wetting as compared with first-order wetting \cite{RMBIB1996,RBBM2004}.

\end{document}